\documentstyle[11pt]{article}

\newdimen\paperwidth
\newdimen\paperlength
\newdimen\margin
\newdimen\vmargin
\textwidth=\paperwidth
\advance \textwidth by -2\margin
\advance\hoffset by -1truein
\advance\hoffset by \margin
\textheight=\paperlength
\advance \textheight by -2\vmargin
\advance\voffset by -1truein
\advance\voffset by \vmargin
\advance\textheight by -2\baselineskip
\begin{document}

\renewcommand{\theequation}{\thesection.\arabic{equation}}
\newcommand{\Section}[1]{\section{#1}\setcounter{equation}{0}}

\begin{titlepage}
\title{ {\bf Strongly Interacting Spinless Fermions in D=1
and 2 Dimensions:}\\
 {\bf A Perturbative-Variational Approach  }\thanks{e-mail:
mardel@fis.ucm.es (m.a.m-d), sierra@cc.csic.es (g.s.)}}

\vspace{2cm}
\author{ {\bf Miguel A. Mart\'{\i}n-Delgado}\dag \mbox{$\:$}
and {\bf Germ\'an Sierra}\ddag \\
\mbox{}    \\
\dag{\em Departamento de F\'{\i}sica Te\'orica I}\\ {\em
Universidad Complutense 28040-Madrid, Spain } \\
\ddag{\em Instituto de Matem\'aticas y F\'{\i}sica
Fundamental}\\ {\em CSIC, 28006-Madrid, Spain} }
\vspace{5cm}
\date{}
\maketitle
\def\baselinestretch{1.3}
\begin{abstract}

We propose a perturbative-variational approach to
interacting fermion systems  on $1D$ and $2D$ lattices at
half-filling.
 We address relevant issues such as the existence of Long
Range Order,  quantum phase transitions and the evaluation
of ground state energy. In $1D$  our method is capable of
unveiling the existence of a critical point in the  coupling
constant at $(t/U)_c=0.7483$ as in fact occurs in the exact
solution at a  value of $0.5$. In our approach this phase
transition is described  as an example of Bifurcation
Phenomena in the variational computation  of the ground
state energy. In $2D$  the van Hove singularity plays an
essential  role in changing the asymptotic behaviour of the
system for large values of $t/U$.  In particular, the
staggered magnetization for large $t/U$ does not display
the  Hartree-Fock law $(t/U) e^{-2 \pi \sqrt{t/U}}$ but
instead  we find the law
 $(t/U) e^{- \frac{\pi ^2}{3} t/U}$.  Moreover, the system
does not exhibit bifurcation  phenomena and thus we do not
find a critical point separating a CDW state  from a fermion
``liquid" state.

\end{abstract}

\vspace{2cm} PACS numbers: 05.20.-y, 02.90.+p, 75.10.-6

\vskip-21.0cm
\rightline{UCM-CSIC}
\rightline{{\bf February 1995}}
\vskip3in
\end{titlepage}

\newpage

\section{Introduction}

  The study of strongly correlated fermion systems apart
from its theoretical  interest has important  practical
applications ranging from high-$T_c$ superconductivity to
quasi-one  dimensional systems. It is usually emphasized the
difficulties in handling models describing correlated
electrons, such as the Hubbard model. This fact has led to
various types of  approximations schemes, which can be
roughly classified as perturbative, variational,
renormalization group techniques and numerical. The
difficulty of the task suggests perhaps that one should
combine various techniques in order to come as close as
possible to the exact solution,  which is probably imposible
to unveil except in one dimension where there  are models
which happen to be integrable.

   The aim of this paper is to combine perturbative and
variational techniques  in the study of strongly correlated
fermions. We want to apply to these systems  the ideas
developed in reference \cite{estegerman} which were applied
to the
 problem of the Ising model in a transverse field.

   Let us suppose that we are given a hamiltonian of the
form
$H(\lambda ) = H_0 + \lambda H_1$, where $H_0$ has a
nondegenerate  ground state $\psi _0$ and $\lambda $ is a
coupling constant. In reference
\cite{estegerman} it was proposed to construct the ground
state
$\psi (\lambda )$ of $H(\lambda )$ as

\begin{equation}
\psi (\lambda ) = \exp (\sum _{n=1}^{\infty} \lambda ^n U_n)
\ \psi _0    \label{1}
\end{equation}

\noindent Solving perturbation theory in $\lambda $ to order
$\nu $ implies  the knowledge of the collection of operators
$\{ U_n \}_{n=1, \ldots ,\nu }$.  Each operator $U_n$
consist in fact of a sum of ``irreducible" operators $V_I$,

\begin{equation}
 U_n = \sum _I p_{n,I}
V_I
\label{2}
\end{equation}

\noindent Hence inserting (\ref{2}) into (\ref{1}) and
interchanging  the order of the sums one arrives to a
``dual" description of the ground state

\begin{equation}
\psi (\lambda ) = \exp (\sum _{I} \alpha _I (\lambda ) V_I)
\ \psi _0    \label{3}
\end{equation}

\noindent where $\alpha _I (\lambda ) = \sum _n \lambda ^n
p_{n,I}$.

  This expression suggests an alternative approximation to
the  ground state $\psi (\lambda )$ which consists in
choosing only a  class of irreducible operators $V_I$  whose
weights $\alpha _I$ are  determined variationally. This was
precisely the approach applied in
\cite{estegerman} to the Ising model in a transverse field,
and it is  our purpose to apply it in this paper to a system
of spinless  fermions defined on a hypercubic lattice in $D$
dimensions.  The hamiltonian for this system is given by:

\begin{equation} H = -t \sum _{<i,j>} (c_i^{\dagger } c_j +
c_j^{\dagger } c_i )   +
\sum _{<i,j>} U (n_i - \frac{1}{2}) (n_j -
\frac{1}{2})                                   \label{4}
\end{equation}

\noindent where $c_i (c_i^{\dagger })$ are fermion
annihilation  (creation) operators satisfying standard
canonical anticommutation  relations (CAR algebra), $n_i =
c_i^{\dagger } c_i$ is the number  operator, $t$ is the
hopping parameter and $U>0$ is a repulsive  coupling
constant mimicking the residual Coulomb interaction.  The
sum in (\ref{4}) is extended over all the links of the
lattice.  The lattice we shall be dealing with is made up of
$L$ sites in  each spatial direction. This amounts to a
total number of sites
$N = L^D$ and it is taken to be even for simplicity. The
hopping  parameter is defined positive $t>0$ and Periodic
Boundary  Conditions are assumed in each direction.

 We shall be concerned  with the study of the ground state
of $H$  when the system is at {\em half-filling}, i.e., when
the number of  electrons $N_e$ is half the number $N$ of
lattice sites
$N_e = <\sum _i n_i> = N/2$ (so that by expanding (\ref{4})
it means that the chemical potential is fixed by $\mu = D
U$),  and secondly when the system is in the strong coupling
r egime, i.e., $U>t$. This means in particular that our
unperturbed
 hamiltonian is

\begin{equation} H_0 = \sum _{<i,j>} U (n_i - \frac{1}{2})
(n_j - \frac{1}{2})                    \label{4b}
\end{equation}

\noindent For $U>0$ and at half-filling there are two ground
states  of $H_0$ corresponding to the unperturbed charge
density states
$ |CDW (e/o) \rangle$,

\begin{equation} | CDW (e/o)>  = \prod _{i \in \Lambda
_{e/o}} c_i^{\dagger} |0>              \label{5}
\end{equation}

\noindent where we have taken into account that the
hypercubic lattice  with $N$ even can be divided into two
interpenetrating sublattices
$\Lambda _e$ ``even" and $\Lambda _o$ ``odd" (or black and
blank sites).

Even though $H_0$ has two different ground states we shall
apply our  technique to each of the CDW states (\ref{5}).
This eventually means that
 after switching on the hopping term in (\ref{4}) each
initial CDW evolve into different CDW-states $\psi _{e/o}$,
characterized by different values of the staggered
magnetization $m_{st}$,

\begin{equation} m_{st} = \frac{2}{N}  <\sum _i (-1)^i
n_i>                                              \label{6}
\end{equation}

\noindent where $(-1)^{i}=1 (-1)$ for $i \in \Lambda _e
(\Lambda _o)$.  For the initial CDW states take on the
values $m_{st}(even/odd) = (+1/-1)$.

An interesting question arises as to whether $m_{st}$
vanishes for a  critical value of $(t/U)_c$ different from
infinity. Equivalently this  amounts to the existence or not
of Long Range Order (LRO) in the  fermion system at zero
temperature. It is clear that if $t/U=\infty$ the  ground
state of (\ref{4}) is a Fermi-sea of free particles which
has no
 preferred distribution in the space which in turn implies
vanishing  staggered magnetization $m_{st}=0$. In
1-dimension the model
 (\ref{4}) can be solved exactly \cite{yang} and one obtains
as the  critical point the value $(t/U)_c=1/2$. As a matter
of fact, the  hamiltonian (\ref{4}) is equivalent to the
anisotropic $XXZ$-Heisenberg  hamiltonian, namely,

\begin{equation} H_{XXZ} = -\frac{1}{2} \sum _{i=1}^N
(\sigma _i^x \sigma _{i+1}^x +
\sigma _i^y \sigma _{i+1}^y + \Delta \sigma _i^z \sigma
_{i+1}^z)            \label{7}
\end{equation}

\noindent This mapping is carried out by the standard
Wigner-Jordan
 transformation and thus the relation between the anisotropy
parameter $\Delta $ and those appearing in (\ref{4}) is
given by,

\begin{equation}
\Delta = -\frac{1}{2
(t/U)}
\label{8}
\end{equation}

\noindent The value $\Delta = -1$ (i.e. $(t/U)_c=1/2$)
corresponds  precisely to the Isotropic Antiferromagnetic
Heisenberg model in
 (\ref{7}). The unperturbed $|CDW (e/o)\rangle$ states of
the fermion
 model are mapped onto Neel states in the Heisenberg model.
We must recall here that mean field theories or Hartree-Fock
approximations to (\ref{4}) always yield a non-vanishing
staggered  magnetization $m_{st}$ no matter how small is the
coupling  constant $U$ or the dimension of the lattice.
 This result is clearly wrong as we have already pointed out
above  for $D=1$ and is explained on the basis that the HF
method does  not take properly into account the quantum
fluctuations which are  very strong in $1D$.  In reference
\cite{shankar} this issue has been analized using
RG-techniques showing that the existence of a solution of
the gap  equation for the CDW order parameter can be traced
back to the fact  that in perturbation theory one is
disregarding some diagrams (those  of BCS-type according to
the terminology of reference \cite{shankar}),  and when they
are properly taken into account they balance the
contributions from the diagrams which favor the
CDW-instabilities.  The outcome of this one-loop
RG-calculation in $D=1$ is a zero beta  function $\beta (U)
= 0$ for the coupling constant $U$. This result is  by all
means only valid in the perturbative regime (i.e. U/t small)
and  implies the existence of a non-trivial fixed point of
the Renormalization  Group which correspond to a Luttinger
liquid. Application of the same  technique to the case of
higher dimensions $D>1$ always encounters  a CDW-instability.

In summary we have the following situations,

\begin{itemize}

\item $(t/U)_c = 1/2$  in $D=1$. Exact.

\item $(t/U)_c = \infty$  in Mean Field Theory, $\forall D$.

\item $\beta (U) = 0$  in Renormalization Group Approach,
$D=1$.

\end{itemize}

\noindent We shall address in detail the issue of the
critical  value $(t/U)_c$ in the system of interacting
fermions as well  as other problems that emerge in
connection to this one.

\section{The Perturbative-Variational Ansatz for Interacting
Fermions}

According to the perturbative-variational (PV) method
developed in  reference \cite{estegerman} we must first
determine the form of the  set operators $\{ U_n \}$ by
inserting the exponential ansatz  (\ref{1}) in the
Schrodinger equation for $H$. Then these operators  serve us
to construct variational wave functions by inserting them
back in the exponential ansatz. The perturbative equations
that  determine the $U_n$ for the lowest order in
perturbation theory  obey the equations,

\begin{equation}
\mbox{order $\lambda$:} \ \ \ \ ([H_0,U_1] + H_1) \psi _0 =
E^{(1)} \psi _0         \label{8b}
\end{equation}

\begin{equation}
\mbox{order $\lambda ^2$:} \ \ \ \ ([H_0,U_2] + [H_1,U_1] +
1/2 ([[H_0,U_1],U_1])) \psi _0 = E^{(2)} \psi
_0                                         \label{8c}
\end{equation}

\noindent where $E^{(1)}$ and $E^{(2)}$ are the perturbative
energies to first and second order.

\noindent A solution of equations (\ref{8b}) and (\ref{8c})
in the  case where $H_0$ is given by equation  (\ref{4b})
and $H_1$ is the hopping term of (\ref{4}) is the following,

\begin{equation} U_1 = \frac{1}{2D-1}\sum _{<i,j>}
T_{ij}                                    \label{9}
\end{equation}

\begin{equation} U_2 = \sum _{<i,j;k,l>}  C_{ijkl} T_{ij}
T_{kl}                                  \label{9b}
\end{equation}

\noindent where $T_{ij}=c_i^{\dagger } c_j+c_j^{\dagger }
c_i$ and
$<i,j;k,l>$ is a path on the lattice from the point $i$ to
the point
$l$ without $i\neq j \neq k \neq l$ and $C_{ijkl} $ some
constants  which may depend on the shape of the path and on
the dimension
$D$, whose precise value is not important for our
discussion.  The perturbative energy up to second order is

\begin{equation} E = -\frac{N U D}{4} (1 + \frac{4}{2D-1}
(\frac{t}{U})^2)                                 \label{9c}
\end{equation}

\noindent Observe that $U_1$ is proportional to the kinetic
term  in the original hamiltonian $H$. The solution of these
equations  is by no means unique. We can add to $U_2$ an
operator of the  form $T_{il}$ with $i,l$ on the same
sublattice. The action of such  an operator on the state $|
CDW \rangle$ is zero.

Now the second stage of the method, according to eqs.
(\ref{1})-(\ref{3}),
 consists in exponentiating the $U_n$ operators (\ref{9})
and (\ref{9b}).  The simplest ansatz takes into account only
the $U_1$ operator yielding

\begin{equation}
\psi _{e/o}^{(1)} (\alpha ) = \exp (\frac{\alpha }{2} \sum
_{<i,j>} (c_i^{\dagger} c_j + c_j^{\dagger} c_i)
)                                   |CDW \
(e/o)>
\label{10}
\end{equation}

\noindent whereas if we take into account the second order
operator
 $U_2$ the ansatz will become

\begin{equation}
\psi _{e/o}^{(2)} (\alpha ) = \exp (\frac{\alpha }{2} \sum
_{<i,j>} T_{ij} + \frac{1}{2}
\sum _{<ij;kl>} \alpha _{ijkl} T_{ij} T_{kl})  |CDW \
(e/o)>                             \label{10b}
\end{equation}

\noindent where $ \alpha _{ijkl} $  will only depend on the
shape of a
$4$-step non-backtracking path. In both cases the parameters
$\alpha ' s$
 are used to minimize the energy of the trial states, which
fixes their
 dependence on the ratio $t/U$. For $t/U$ small we must have
$\alpha \sim t/U$, $\alpha _{ijkl} \sim (t/U)^2$ and such
that
$\psi ^{(1)}$ ($\psi ^{(2)}$) agrees to first (second) order
with  the exact ground state of $H$.

\noindent We notice that this is the essence of the PV
method  proposed in \cite{estegerman}, namely, to use the
results obtained  to a given order in perturbation theory in
order to establish a  variational calculation which extends
the validity of the perturbative  results beyond
perturbation theory.

\noindent In what follows we shall restrict ourselves to the
analysis  of the first order ansatz $\psi ^{(1)} (\alpha )$
(hereafter called $\psi  (\alpha )$).

\noindent As $t/U$ increases we expect also $\alpha $ to
increase,  after all the role of this hopping term in
(\ref{4}) is to  disorder  the  CDW state, but that is
precisely what the exponential operator in  the ansatz
(\ref{10}) is doing on the CDW-states.

\noindent Eventually, as $t/U$ increases we may get $\alpha
= \infty$  in which case, as is shown below, the state
(\ref{10}) becomes the  Fermi-sea or ground state of $H_1$,

\begin{equation}
\lim_{\alpha \rightarrow \pm \infty} \psi _{e/o} (\alpha )
\sim    |\mbox{Fermi Sea}>  = \prod _{k \in Fermi \ Sea}
c^{\dagger}_k |0>           \label{11}
\end{equation}

\noindent All the states $\psi (\alpha )$ with $\alpha $
finite have  non-vanishing staggered magnetization and hence
only for $\alpha = \infty$  we get a state with $m_{st}=0$.
This in turn implies that the value of
$t/U$ at which $\alpha = \infty$ is nothing else but the
critical value
$(t/U)_c$. Therefore the rule to find in our approach the
critical value
$(t/U)_c$ is given by finding the minimum of the variational
energy
$E(\alpha ;t/U)$ of the state $\psi (\alpha )$ for $t/U =
const.$.  This fixes $t/U$ as a function of $\alpha $, i.e.,
$t/U = \cal {F} (\alpha )$.  Then the critical value
$(t/U)_c$ at which the staggered magnetization  vanish is
given by

\begin{equation} (t/U)_c = \lim_{\alpha \rightarrow \infty}
\cal {F}(\alpha)        \label{12}
\end{equation}

\noindent If $(t/U)_c$ turns out to be non-zero then the
ansatz (\ref{10})  would be valid in the whole range $0\leq
t/U \leq (t/U)_c$ and beyond  this the variational state
remains the free Fermi sea (\ref{11}). Hence  to get more
information we should study the model from the ``other"
side of coupling constant, as for example perturbation
theory in $U$  or some other method.

To prove (\ref{11}) notice that the ground state of the
operator appearing
 in the exponential of (\ref{10}) is, for either sign of
$\alpha $, given by the Fermi sea at half-filling. Now if we
let $\alpha $  go to $\infty $, and we
 introduce the resolution of the identity we project the
state $\psi (\alpha )$
 onto $|Fermi \ Sea \rangle$. We note that a similar
technique is used in the so called projected Monte Carlo
method in order to filter  out the ground state, \cite{promc}

\begin{equation}
\lim_{t\rightarrow \infty} e^{-t H} |\phi > \sim  |G.S. \ of
\ H> <G.S. \ of \ H|\phi >           \label{13}
\end{equation}

\noindent It is assumed that the trial state $|\phi >$ is
not orthogonal
 to the true ground state of $H$.

\noindent For the ansatz (\ref{10}) we do not have to resort
to numerical  computations to obtain the mean value of $H$
since this state is in
 fact a Hartree-Fock state of a restricted type.

\noindent To prove this we shall express the state
(\ref{10}) in momentum  space. For that purpose we introduce
the creation operator $c_k^{\dagger }$
 and annihilation operator $c_k$ which are the Fourier
transformed of $c_i$,

\begin{equation} c_k^{\dagger } = \frac{1}{\sqrt{N}} \sum _i
e^{i k \cdot R_i}  c_i^{\dagger }          \label{14}
\end{equation}

\noindent where $k = \frac{2 \pi }{L} (n_1, \ldots ,n_D)$,
with $n_i = 0,1, \ldots ,L-1 \ mod(L)$.

\noindent The CDW  states (\ref{5}) can then be  expressed
using (\ref{14}) as,

\begin{equation} |CDW (e/o)>  =  \frac{1}{\sqrt{2}}\prod _k
' (c_k^{\dagger } \pm c^{\dagger }_{k+Q})|0>
\label{15}
\end{equation}

\noindent where $Q = (\pi , \ldots ,\pi )$ is the AF vector
and $\prod _k^{'} $ stands for the product over the reduced
Brillouin zone, also called magnetic zone.

\noindent Taking into account that:

\begin{equation} -\sum _{<i,j>} (c_i^{\dagger } c_j +
c_j^{\dagger } c_i)  = \sum _k \epsilon _k n_k    \label{16}
\end{equation}

\begin{equation}
 \epsilon _k = -2 \sum _{a=1}^D \cos
(k_a)                               \label{16b}
\end{equation}

\noindent and  $n_k = c_k^{\dagger } c_k$, we can write the
states (\ref{10}) as

\begin{equation}
\psi _{e/o} (\alpha ) = \frac{1}{\sqrt{2}} \prod '_k
e^{-\frac{\alpha }{2} \epsilon _k} c_k^{\dagger } \pm
 e^{-\frac{\alpha }{2} \epsilon _{k+Q}} c_{k+Q}^{\dagger } )
|0>                                    \label{17}
\end{equation}

\noindent From this expression we can inmediately derive
equation (\ref{11}).

Concerning this result we would like to make some comments:

\begin{description}
\item[i)] Equation (\ref{17}) shows that the PV ansatz
(\ref{10})  is a Hartree-Fock state, i.e., it is a state of
the form

\begin{equation}
\psi _{HF} = \prod _k ' (u_k c_k^{\dagger } +
 v_k c_{k+Q}^{\dagger } )
|0>                                        \label{18}
\end{equation}

\noindent with

\begin{equation}
\begin{array}{l} u_k = \frac{1}{\sqrt{2}}e^{-\frac{\alpha
}{2} \epsilon _k} \\
                                         \\ v_k =
\frac{1}{\sqrt{2}} e^{-\frac{\alpha }{2} \epsilon
_{k+Q}}
\end{array}
\label{18b}
\end{equation}

\item[ii)] Since the functional dependence of $u_k$ and
$v_k$  is fixed by equation (\ref{18b})  he PV state is a
{\em restricted} Hartree-Fock state. This implies  that the
variational method applied to the PV state using $\alpha $
as the unique variational parameter, needs not coincide with
the  variational method applied to an unrestricted HF state
where all the
 parameters $u_k$, $v_k$ are variational \cite{rr},
\cite{ogu}.

\end{description}

The computation of the mean value of the hamiltonian $H$
(\ref{4})  on the state (\ref{17}) is standard, we quote
here the result and  leave the details of the proof for the
appendix A,

\begin{equation} E \equiv -\frac{1}{4} N D U \ e(t/U,\alpha
)                              \label{19}
\end{equation}

\noindent where $e(t/U,\alpha ) $ is the reduced energy

\begin{equation} e(t/U,\alpha ) = 4 \frac{t}{U}  I_1(\alpha
)  + I_0^2(\alpha ) +  I_1^2(\alpha )
\label{19b}
\end{equation}

\noindent and $I_0(\alpha )$, $I_1(\alpha )$ are the
following integrals,

\begin{equation} I_0(\alpha ) = \int _{\Omega } \frac{d^D
k}{(2 \pi)^D}
 \frac{1}{\cosh (\alpha \epsilon _k)}   \label{20}
\end{equation}

\begin{equation} I_1(\alpha ) = \frac{1}{2 D} \int _{\Omega
} \frac{d^D k}{(2 \pi)^D}
\epsilon _k \tanh (\alpha \epsilon
_k)                                    \label{20b}
\end{equation}

\noindent Here $\Omega $ denotes the interior of the
Brilluoin zone  ($-\pi \leq k_a \leq \pi $).

\noindent Minimization of $E$ with respect to $\alpha $
yields

\begin{equation} t/U = -\frac{1}{2} (I_1 + \frac{I_0
I_0^{'}}{I_1^{'}})
\label{21}
\end{equation}

\noindent where $I_0^{'}=dI_0/d\alpha $, etc. For a given
value of $t/U$  there may exist several values of $\alpha $
satisfying (\ref{21}).  The desired value is selected by
minimizing the energy $E$.  This equation will play the role
for us of the standard gap  equation of the HF method.

 We can perform a consistency check of our formalism which
consists in verifying that for small  parameter $\alpha $ it
is  linearly related to the effective coupling constant
$t/U$.  This must be  so because it means that we are very
close  to the ground state of $H_0$, the CDW state, and thus
we  are in the perturbative regime where both quantities are
proportional, for that is the origin of the construction of
the
$U_1$ operator (\ref{9}).

\noindent  To prove this assertion we must evaluate the
 leading behaviour when $\alpha \rightarrow 0$ of the basic
integrals $I_0$, $I_1$ in (\ref{20}), (\ref{20b}). From
these expressions we arrive at,

\begin{equation}
\begin{array}{ll}  I_0 \stackrel{\alpha \rightarrow 0}{\sim
} 1 &  I_0' \stackrel{\alpha \rightarrow 0}{\sim }  -2 D
\alpha \\
                &                                \\ I_1
\stackrel{\alpha \rightarrow 0}{\sim } \alpha  &  I_1'
\stackrel{\alpha \rightarrow 0}{\sim }  1
\end{array}
\label{22}
\end{equation}

\noindent Inserting these small-$\alpha $ behaviours
 in the defining relation (\ref{21}) of $t/U$ in terms of
$\alpha $ we find

\begin{equation} t/U \simeq  \alpha (D -
\frac{1}{2})
\label{23}
\end{equation}

\noindent which is indeed the linear relation previously
advanced in equation (\ref{9}). This relationship is made
explicit in Figure 2
 where we have plotted the corresponding equation (\ref{23})
for
 the cases $D=1$ and $2$ dimensions.

Once this check is done the big issue is to see whether we
can  extract nonperturbative results for the system of
interacting fermions  by going over large values of  $\alpha
$ for which the perturbative  relation (\ref{23}) ceases to
be true. In the forthcoming sections  we investigate this
issue in one and two dimensions.


\section{Strongly Interacting Fermions in $D=1$ and the
Heisenberg Model}


Let us consider the system of spinless fermions interacting
 by means of the hamiltonian $H$ (\ref{4}) in one
dimensions.  This is a very special case where exact
formulas exist for the
 magnitudes of interest: ground state energy, magnetization
or  Long Range Order, density correlators, etc $\ldots $.
This allows  us to confront our PV methods predictions with
the exact results. Moreover,  this is a very demanding test
ground for it is well-known that mean  field theory
treatments are usually doomed to failure because the  strong
quantum fluctuations occuring in one dimensions are not
properly taken into account.

In one dimension the system undergoes a quantum phase
transition at the critical value of the coupling constant
$(t/U)_c=1/2$.  This is an essential singularity phase
transition.  For small  values of $t/U$ the system is in an
ordered phase  in coordinate space corresponding to a
perturbation of the  CDW state. In this phase the staggered
magnetization calculated  by Baxter \cite{baxter} is
non-vanishing. For large values of $t/U$ the  system is in a
disordered phase similar to a  fermion ``liquid" of
Luttinger type.

Elucidating the  existence of a finite critical value
$(t/U)_c$
 in this case with our method goes over the determination of
the  asymptotic behavior of integrals $I_0$, $I_1$ and its
derivatives  for $\alpha \rightarrow \infty $.

Let us outline how to compute the leading large-$\alpha $
behaviour
 of $I_0^{(D=1)}$ for it reveals the role played by the
Fermi points  in the physics of the system, and  it is also
useful to generalize the  same technique to the more
involved case of two dimension discussed  in next section.

\noindent For $\alpha \rightarrow \infty $ only the regions
in the  neighborhood of the points satisfying $\cos k =0$,
that is the  Fermi points $k = \pi /2$ and $- \pi /2$,
contribute significantly  to the integral $I_0$ in equation
(\ref{20}) for $D=1$.  As matter of fact, both Fermi points
contribute the same  amount to the integral under
consideration. Thus, it is convenient  to introduce a
cut-off $\Lambda $ to isolate the contribution around  the
point, say $k=\pi /2$:

\[ I_0^{(D=1)} \sim 2 \int _{\frac{\pi }{2}-\Lambda
}^{\frac{\pi }{2}+\Lambda } \frac{d k}{2 \pi}
\frac{1}{\cosh (2 \alpha \cos k)}
\]

\noindent Shifting variables $k=\pi /2 + \kappa $ with
$\kappa $
 small and thereby approximating $\cos k \simeq -\kappa $,
we may write

\[  I_0^{(D=1)} \sim \frac{2}{2 \pi} \int _{-\Lambda
}^{\Lambda } d \kappa \frac{1}{\cosh (2 \alpha \kappa )}
\]

\noindent Changing variables again $x=2 \alpha \kappa $ we
can  freely take the limit $\alpha \rightarrow \infty $ in
the limits of  integrations and arrive at

\[ I_0^{(D=1)} \sim \frac{1}{2 \pi \alpha } \int _{-\infty
}^{\infty } \frac{d x}{\cosh x} = \frac{1}{2 \alpha}
\]

\noindent This result establishes that $I_0^{(D=1)}$
vanishes  as a power of $\alpha $ with exponent $-1$ in the
asymptotic  region $\alpha \rightarrow \infty $. It is
straightforward to extend  this analysis to the other
integrals $I_0'$, $I_1$ and $I_1'$ so that we
 summarize the results hereby:

\begin{equation}
 \begin{array}{ll} I_0^{(D=1)} \sim \frac{1}{2 \alpha }  \
\  & I_0^{(D=1)'} \sim -\frac{1}{2 \alpha ^2}   \\
       &                             \\ I_1^{(D=1)} \sim
\frac{2}{\pi} \ \   & I_1^{(D=1)'} \sim \frac{\pi }{24
\alpha ^3}
\end{array}
\label{24}
\end{equation}

\noindent Now, in order to determine the critical value
$(t/U)_c$
 as prescribed in equation (\ref{12})
 we must insert the leading values  (\ref{24}) in the
defining  variational relation of $t/U$ (\ref{21}) This
inmediately yields the critical value
$(t/U)_c=\frac{2}{\pi}\simeq 0.63662$  which is quite close
to the exact value  $(t/U)_c=0.5$.

\noindent However this quick derivation is not correct. The
first  indication that something peculiar is going on in one
dimensions  comes from the shape of $t/U$ as a function of
$\alpha $ in  equation (\ref{21}). This function has the
horizontal asintota $2/\pi $ as
 explained above but it is not a monotonous increasing
function.  In fact, it exhibits a local maximum at $\alpha =
2.3307$. This in  turn means that the relationship
(\ref{21}) is not invertible so that we  cannot express
$\alpha $ as a function of $t/U$ to be inserted in the
energy equation and the remaining formulas.

\noindent Therefore, to determine the correct critical value
$(t/U)_c$
 we must proceed more carefully and in doing so we shall
unveil the  peculiarities of the PV method when applied to
the special case of  one dimensions. The correct procedure
is exemplified in Figure 1  where we plot the reduced energy
$e^{(D=1)} (\alpha ;t/U)$ as a  function of $\alpha $ for
varying values of the coupling constant $t/U$  which is
considered as a parameter. In this fashion we are correctly
implementing the minimization program thereby assuring that
we are  picking up the absolute maximum of $e^{(D=1)}
(\alpha ;t/U)$
 (recall that a maximum in the reduced energy correspond to
a  minimum of  the ground state energy $E$ (\ref{19}) ), not
just the  local maximum. This local maximum contribution was
the one we  were only considering previously when the
critical value $2/\pi $ was
 naively obtained.

\noindent For small values of $t/U$, say $0.2$ (see Figure
1a),  the function $e^{(D=1)} (\alpha ;t/U)$ exhibits an
absolute maximum
 for a finite value of $\alpha $ and an absolute minimum for
 $\alpha \rightarrow \infty$.

\noindent When $t/U$ is increased, say $0.4$ (see Figure
1b),  the relative distance between the maximum and the
minimum is reduced.

\noindent Eventually, for say $t/U \sim 0.735$ (see Figure
1c), a local  minimum shows up nearby the absolute maximum
and the minimum  at $\alpha \rightarrow \infty $ becomes  a
local maximum. Yet, the first  maximum dominates over the
new maximum at infinity and it is thereby  picked up in our
minimization procedure.

\noindent We are able to compute the exact value taken by
 $e^{(D=1)} (\alpha ;t/U)$ in the $\infty $, as a function
of $t/U$.  In fact, substituting the asymptotic behaviours
 (\ref{24}) in $e^{(D=1)} (\alpha ;t/U)$ we get,

\begin{equation}
 e^{(D=1)} (\infty ;t/U) = \frac{8}{\pi }  \frac{t}{U} +
(\frac{2}{\pi })^2  \label{25}
\end{equation}

\noindent This is an increasing function of $t/U$ and
eventually  this local maximum will overcome the thus far
absolute
 maximum occurring at a finite $\alpha $. When this fact
happens, we are obtaining the true value of the critical
point $(t/U)_c$.

\noindent As the coupling constant $t/U$ keeps being raised,
 two things happens: the local minimum moves towards the yet
absolute maximum and the relative height of the maximums is
reduced. The critical point occurs when the height of the
local  maximum at $\alpha $ infinity equals that of the
maximum occurring  at finite $\alpha $. This happens for the
following value

\begin{equation} (t/U)_c = 0.748375 \ \ \ \ \ \ (\alpha _c =
1.80032)  \label{26}
\end{equation}

\noindent From this value of $t/U$  the maximum at infinity
becomes
 the absolute maximum henceforth.

\noindent It is worth pointing out that in our procedure for
one dimensions  explained above we have been implicitly
describing an example  of what has come to be known {\em
Bifurcation Phenomena} in  several branches of Physics (We
shall see in next section that
 such phenomenon is absent in two dimensions.)

\noindent As a matter of fact, this way of thinking is well
suited for
 our problem as well as enlightning. Bifurcation phenomena
typically  occurs in a system subject to the action of a
certain potential say
$V(x;p)$  where $x$ can be a generalized coordinate
describing the
 state of the system and $p$ an external parameter which is
somehow varied. For a certain initial value of $p$, the
potential
$V$ exhibits two minima $x_{m1}<x_{m2}$ with  $V(x_{m1}) <
V(x_{m2})$  and the system locates at $x_{m1}$. As the
external parameter $p$ is varied,  the two minima start
approaching each other till eventually  they meet,
$V(x_{m1})=V(x_{m2})$. This in turn defines the  critical
value $p_c$. As $p$ is yet more increased, the two minima
exchange their roles, $V(x_{m1}) > V(x_{m2})$, and the
system  locates at $x_{m2}$. Altogether, when $p$ is varied
at one go  over its full range of validity, the system
undergoes a ``phase transition" from
 the state at $x_{m1}$ to the state at $x_{m2}$. But this is
precisely
 what we have been describing above in our method where
$e^{(D=1)} (\alpha ;t/U)$ plays the role of the potential
function
$V(x;p)$, $\alpha \sim x$ and $t/U \sim p$. Interestingly
enough, the phase transition we are describing  in this
fashion is the quantum phase transition occurring in the
spinless interacting fermion system from the CDW state to
the  Fermi sea state.

\noindent This result is quite different from what one
obtains using  mean field theory, where there is always a
solution of the gap  equation for $\Delta _{CDW}$ no matter
how small is the coupling  constant $U$. The physical
explanation of this result is that any perturbation
 having momentum $\pi $ links the ground state, given by the
Fermi sea at half-filling, to states of arbitrarily low
energy in which
 a particle just below right (left) Fermi point is pushed
above the
 left (right) Fermi point. In this way the Fermi sea at
half-filling  becomes unstable under this perturbation
turning into a CDW state. Our result above in $D=1$ is
precisely the opposite. It shows that  starting from the
CDW-ground-state it becomes unstable for $t/U$  bigger than
a critical value $(t/U)_c$.

We can summarize our discussion in the following diagram,
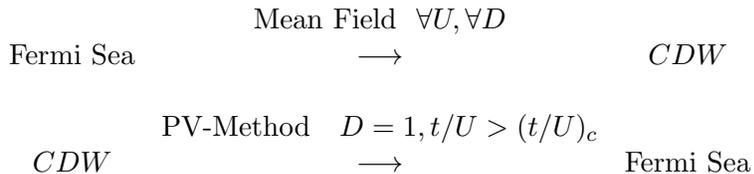
\[
\begin{array}{ccc}
      &  \mbox{Mean Field}  \ \ \forall U, \forall D &  \\
\mbox{Fermi Sea} & \longrightarrow & CDW \\
      &    &     \\
      & \mbox{PV-Method} \ \ \  D=1, t/U > (t/U)_c    &  \\
CDW  &  \longrightarrow  & \mbox{Fermi Sea}
\end{array}
\]

\noindent Of course the correct result is that the CDW
state for $t/U > 1/2$ should become a Luttinger liquid,  but
it is clear that in our approach we have not introduced
enough parameters to tell the difference between a free
Fermi  liquid and a Luttinger liquid.

The outcome of the procedure oulined in Figure 1 is to
achieve the correct set of values $(t/U,\alpha )$ associated
 to the PV method. When this collection of values is plotted
it looks like in Figure 2. Notice firstly that $t/U$ as a
function  of $\alpha $ is single-valued and secondly, that
from a finite  initial value $\alpha _c=1.80032$ the
critical value $(t/U)_c=0.748375$ is  reached once and for
all.

Let us now turn our attention to physical quantities and
consider first the staggered magnetization $m_{st}$
(\ref{6})  or equivalently the averaged fermion density
depending on
 whether we choose the anisotropic Heisenberg viewpoint
(\ref{7})  or the spinless fermion description (\ref{4}).

\noindent First of all, notice that the staggered
magnetization
 $m_{st}$ (\ref{6}) precisely coincides in our method with
the
 basic integral $I_0(\alpha )$ (\ref{20})

\begin{equation} m_{st} (t/U )=
I_0(t/U)                                       \label{27}
\end{equation}

\noindent In fact, by going to momentum space and using the
 results of Appendix A we find

\[ m_{st} = \frac{2}{N} \sum _i (-1)^i <n_i> = \frac{2}{N}
\sum _k <c_k^{\dagger } c_{k+Q}> =
\sum _k \frac{1}{\cosh (\alpha \epsilon _k)}
\]

\noindent However, in order to extract the physics underlying
 this magnitude we must not plot $I_0$ as a function of
$\alpha $
 but instead as a function of the coupling constant $t/U$.
After all, the parameter $\alpha $ is somehow an artifact
of our method which is related to the physical parameter
$t/U$ by the plot in  Figure 2. Expressing $\alpha $ as a
function of $t/U$ in $I_0(\alpha )$ we obtain the plot of
the  magnetization $m_{st}$ which we show in Figure 3. In
this figure  we also plot the exact magnetization as given
by Baxter \cite{baxter}.  Several remarks are in order. Only
for small values of $t/U$ there
 is a good quantitative agreement between both methods.
Beyond these values $t/U \sim 0.2$ there is major
quantitative  disagreemet. Yet, our approximate method gives
a  reasonable qualitative behaviour in the sense that it
predicts  the existence of a critical value $(t/U)_c$ beyond
which the  magnetization totally vanishes. However the PV
method  (to lowest order) predict an otherwise absent finite
jump in  the magnetization at the critical value.

In all, the physical meaning of the magnetization graphics
 is that we can predict with our method the absence of  LRO
for small $t/U$ at zero temperature for the system  of
spinless interacting fermions.

Now we move on to consider another relevant physical
quantity such as the ground state energy (\ref{19}) as a
function of $t/U$.  In Figure 4 we plot the reduced energy
 (\ref{19b}) as obtained by the PV method against the exact
result given by several authors \cite{orbach},
\cite{walker}.  The agreement is considerably much better
than with the  magnetization. In fact there is a very good
quantitative  agreement  not only for small $t/U$ but for
the whole  physical range till $t/U=0.5$. The major
difference  though is the prediction of a bigger $(t/U)_c$
than the exact result.

\noindent As a precise quantitative check of our values  we
have collected our PV result for the reduced energy  in the
Isotropic Heisenberg Antiferromagnet $(t/U=1/2)$ in  Table 1
along with the exact result \cite{hulthen} and several
other  methods. The result is a good estimation of the
ground  state energy.

\noindent Furthermore, we are also able to compute exactly
within our approximate method any multi-density correlators.
For instance, the density-density correlator $<(n_i-1/2)
(n_{i+1}-1/2)>$  computed in Appendix A has the
interpretation of the  Short Range Order in the Heisenberg
model for it is equivalent  to the spin-spin correlator
$<S_i^z S_{i+1}^z>$. In Appendix A  we show how to compute
it as a function of $t/U$. We do not show  the plot of this
quantity against the exact value but only remark  that there
is a good quantitative and qualitative agreement between
them.

Finally, we close the analysis of the $D=1$ dimension case by
 pointing out that the reduced energy $e(t/U)$ exhibits a
finite  jump in its derivative at the critical value
$(t/U)_c=0.748375$.  The constant value of the derivative at
the right hand side of the  critical point comes from the
linear contribution in equation  (\ref{25}) of the absolute
maximum at $\alpha \rightarrow \infty $  previously
discussed. The meaning of this fact is that our PV  method
predicts a first order quantum phase transition for the
fermion system.

\newpage

\section{ Strongly Interacting Fermions in $D=2$ Dimensions}


After studying the pros and cons of our PV approach to
one-dimensional
 interacting fermions, we address the much more unknown case
of $D=2$  spinless fermions. This case is particularly
interesting in physical applications  such as high-$T_c$
superconductivity.

First of all, we should quote a major difference ocurring
with the $D=2$  fermions: there is {\em no connection} with
the Heisenberg model in 2  dimensions (as far as the present
knowledge is concerned). That is to  say, the Wigner-Jordan
mapping transforming (\ref{4}) onto (\ref{7})  ceases to
work in higher dimensions. Correspondingly, all of our
conclusions will be only valid for the fermion system but
not for the spin system.

The system of $2D$ interacting spinless fermions is worse
understood in comparison with the $2D$ Anisotropic
Heisenberg model.  The issue of LRO has been elucidated for
the spin system in  references \cite{lieb}, \cite{dyson}
using the reflection property  in connection with the
Peierls argument. It turns out that in 2  dimensions there
exists LRO for low temperature $T$ and small  enough
anisotropic parameter. The main obstacle in extending
 similar results to spinless fermions relies on the fact
that no  type of reflection positivity property is known to
hold for fermions.  Recently, though, what has been possible
to prove the most  is the existence of LRO for low but {\em
finite} temperature  and small coupling constant $t/U$
\cite{pirmin}. There is not  exact proof  of this statement
at zero temperature, although it
 is believed that LRO persists for $t/U \ll 1$. In fact,
numerical  methods such as Quantum Montecarlo fails to
provide a definite  answer at zero temperature
\cite{gubernatis}.

\noindent With this considerations in mind the study of
exact  properties for two-dimensional fermions at $T=0$
remains an  open problem and we undertake this study with
our PV method  developed in previous sections.

\noindent We follow a parallel analysis as described in one
dimensions. The first key observation is that when plotting
the function $t/U=t/U(\alpha )$ as given in equation
 (\ref{21}) {\em NO local maximum} shows up as is the  case
in $D=1$. In fact, Figure 2 shows this dependence
 which turns out to be monotonically increasing.  This
property guarantees that the function is invertible  so that
we can trade the $\alpha $ parameter in our PV
 results for the more physical parameter $t/U$.

\noindent Moreover, this conclusion is further backed by a
similar analysis as in Figure 1 for the reduced energy
$e(\alpha ;t/U)$  for increasing values of $t/U$. We have
performed such  analysis with the result that  {\em there is
no Bifurcation Phenomena} present in $D=2$
 spinless fermions. In other words, the function $e(\alpha
;t/U)$
 always ($\forall (t/U)$) has a maximum at finite $\alpha $,
which in turn is the absolute maximum for the whole range
of $\alpha $. The value $e(\infty ;t/U)$ at infinity never
overcomes that maximum neither a local minimum appears  in
the vicinity of that maximum.

\noindent Thus, the minimization procedure of the PV  method
is entirely governed by the maximum at finite
$\alpha $ and the analysis is thereby simplified in $D=2$
dimensions.

The second key observation present in Figure 2 is that
there is not  a finite critical value $(t/U)_c$ of the
coupling  constant towards which $\alpha $ tends
asymptotically at
 infinity. Quite the contrary, $t/U$ always grows as a
function  of $\alpha $. In fact, we can compute the
behaviour  of $(t/U)(\alpha )$ in the asymptotic regime
 $\alpha \rightarrow \infty$ for the $D=2$ case.  To this
purpose, we need to find that behaviour for  the integrals
$I_0(\alpha )$, $I_0 '(\alpha )$, $I_1(\alpha )$ and $I_1
'(\alpha )$
 and then insert them in (\ref{21}).

We have left out the calculation of the leading order
 behaviour of those basic integrals for the Appendix B.  We
hereby present the final results,

\begin{equation}
\begin{array}{cc} I_0^{(D=2)} (\alpha ) \sim  \frac{\ln
\alpha }{2 \pi \alpha } \ \  &
\frac{dI_0^{(D=2)} }{d\alpha } (\alpha ) \sim  -\frac{\ln
\alpha }{2 \pi \alpha ^2}  \\
         &                                           \\
I_1^{(D=2)} (\alpha ) \sim  (\frac{2}{\pi })^2 \ \  &
\frac{dI_1^{(D=2)} }{d\alpha } (\alpha ) \sim  \frac{\ln
\alpha }{24 \alpha ^3}
\end{array}                      \label{29}
\end{equation}

\noindent In doing this calculation it is essential to
 deal with the van Hove singularities (saddle point
singularities)  appearing at the corners $A (\pi ,0)$ and $B
(0,\pi )$  of the Fermi Surface (FS) for two-dimensional
fermions  as in Figure 5. Analogously as in $D=1$
dimensions,  only the region of the Brillouin zone near the
FS contribute  significantly to the integrals when $\alpha
\rightarrow \infty$.  Much care is needed to isolate this
asymptotic behaviour as  we have to distinguish the
contribution of the van Hove points
 from the rest of the FS points which we call regular
 points (see Figure 5). In fact, in the intermediate step,
there is a highly non-trivial cancellation of cut-off
singularities
 between van Hove and regular-points contribution.

\noindent Moreover, we have performed numerical analysis  of
these integrals in the asymptotic region $t/U \rightarrow
\infty$  and we find numerical agreement with the results
expressed  in equation (\ref{29}).

The main feature of equations (\ref{29}) is the presence of
the  logarithmic terms which were absent in $D=1$ dimensions
 (\ref{24}). This is ultimately the signal of the van Hove
singularity  and the reason why $t/U$ grows indefinitely:
substituting equation  (\ref{29}) in $t/U$ (\ref{21}) we get

\begin{equation} t/U \sim -\frac{2}{\pi ^2} + \frac{3}{\pi
^2} \ln \alpha                                  \label{30}
\end{equation}

\noindent which explains the logarithmic growth of $t/U$
present in Figure 2.

\noindent Therefore, our method applied to $D=2$ spinless
fermions does not predict the existence of a quantum phase
transition from the CDW state where we started out to a new
phase, say a disordered fermion ``liquid" state, as was the
case  in $D=1$ dimensions.

The analysis of physical properties such as magnetization
and ground state energy goes in parallel to the $D=1$ case
of section 3. We show our results in Figures 3  and 4.  The
main feature of these plots is the absence of a critical
point $(t/U)_c$ which makes the magnetization decrease  to
zero at infinity and the reduced energy to grow to infinity
on the contrary. In fact, from (\ref{30}) the asymptotic
relation  between $\alpha $ and $t/U$ is

\begin{equation}
 \alpha   \sim e^{\frac{\pi ^2}{3}
t/U}                               \label{31}
\end{equation}

\noindent Upon substitution of (\ref{31}) in the
 magnetization (\ref{27}) we obtain,

\begin{equation} m_{st}   \sim (t/U) e^{-\frac{\pi ^2}{3}
t/U}   , \ \ \ \ t/U \rightarrow
\infty                           \label{32}
\end{equation}

\noindent It is interesting to compare the result (\ref{32})
 with those predicted by a Hartree-Fock method
\cite{hirsch}, which in 1,2 and 3 dimensions are given by

\begin{equation}
\begin{array}{c c c c} m_{st}  & \sim & \frac{t}{U} e^{- 2
\pi \frac{t}{U}} & (D = 1,3) \\
     &   &   &    \\ m_{st}  & \sim & \frac{t}{U} e^{- 2 \pi
\sqrt{\frac{t}{U}}} & (D = 2)  \label{32b}
\end{array}
\end{equation}

\noindent The square root behaviour in the exponential is
due to a combined effect of the nesting of the
$D=2$ Fermi surface at half-filling and the van Hove
singularity of the density of states.
 Hence, the HF result gives a stronger tendency to a  CDW
state in $D=2$ as compared to $D=1$ or $3$.  On the
contrary, the PV method when extrapolated to  the $U$ weak
coupling regime, gives a behaviour similar  to that of
$D=1,3$ HF, the only difference being in the
 factor $\frac{\pi ^2}{3}$ instead of $2 \pi $. It is an
interesting question to discriminate  which of the two
behaviours, namely the square root  law $\sqrt{\frac{t}{U}}$
or the linear
$\frac{t}{U}$ law in the exponent controls the gap in
$D=2$ dimensions at half-filling. We want to stress that
formula (\ref{32}) is obtained  from an extrapolation of the
PV state $\psi (\alpha )$ from  the perturbative region of
the ansatz $(t/U \ll 1)$ to the  non-perturbative $(t/U \gg
1)$ by means of the variational method. We find indirect
 support for this method when taking the $t/U \rightarrow
\infty$  in the PV formulas for we find that the physics  of
the system is governed by the momenta very  close to the
Fermi surface, and the Fermi surface  is a concept which
makes sense for a Fermi sea
 located at $t/U \rightarrow \infty$ in the coupling
constant.

In reference \cite{hirsch} a numerical analysis has  been
done for the $2D$ Hubbard model showing
 that the $e^{-2 \pi t/U}$ is not inconsistent with  the
numerical results.

Let us notice that in the above discussion we are
 comparing asymptotic behaviours concerning the
 staggered magnetization. Another similar analysis
 could be done as far as the gap $\Delta _{CDW}$
 is concerned. In the HF method both quantities are
 easily related by $\Delta _{CDW} =  U m_{st}$, but in our
PV method that relationship is  not that easy to obtain for
one has to go through  the computation of the gap by doing
an extension of
 the PV ansatz to the first excited state of the system.  We
shall not dwell here upon this and other extensions
 which will be reported elsewhere.


\section{Final Considerations}


In section 2 and appendix A we have shown that the
 trial state $\psi ^{(1)}$ based on the lowest order PV
method is equivalent to a  Hartree-Fock state of a
restricted type. An straightforward generalization of the
ansatz (\ref{10}) is given by

\begin{equation}
\psi _{UHF} = \exp (\frac{1}{2} \sum _{<i,j>} c_i^{\dagger }
A_{ij} c_j) \ |CDW>        \label{35}
\end{equation}

\noindent where $A_{ij}$ is an hermitian matrix
 (i.e. $A_{ij}^{\ast } = A_{ji}$). If the matrix $A_{ij}$
is an hermitian matrix only depending on the difference
 $\vec{R_i}-\vec{R_j}$, then it can be shown  that $\psi
_{UHF}$ is an unrestricted HF state  of the form (\ref{18})
where the parameters
$u_k$ and $v_k$ take generic values depending  on the
eigenvalues of the matrix $A$.

Carrying out the PV method to highest orders in
 the coupling constant $t/U$ we find operators as  the ones
in the exponent of (\ref{35}), along with  operators with a
more complicated structure.  This fact may already be seen
from the expression  of the operator $U_2$  (\ref{9b}).
First of all, $U_2$ is a sum of operators involving
$4$-fermion terms, which implies that the  corresponding
state $\psi ^{(2)}$ (\ref{10b}) is {\em not} a HF-state.  In
fact, working with this type of states is not simple at all.
The interesting feature of $\psi ^{(2)}$ is that it contains
 correlation effects which are absent in the  HF-state $\psi
^{(1)}$ (A similar phenomena happens  for the Ising model in
a tranverse field ITF where the  analoge of the $\psi
^{(1)}$ state is also a mean field  state, while the second
order ansatz $\psi ^{(2)}$ is not  mean field and it
incorporates correlations ).

\noindent From a more physical point of view the effects  of
the operators $U_n$ or $V_I$ is that when they act on  the
unperturbed ground state, say $|CDW (even)>$, they  change
locally the even-CDW vacuum into an odd-CDW vacuum.  From
the form of the $U_2$ operator we see that the irreducible
operators $V_I$ can be labelled by clusters of dimers, i.e.,
products  of bilinear operators $T_{ij} =  c_i^{\dagger}
c_j+c_j^{\dagger }c_i$ with the links $<i,j>$ belonging  to
some cluster. The linear clusters are reminescent of the
string operators considered in references \cite{dagotto},
\cite{ss},  but we should in principle consider all possible
clusters in order to
 construct the exact solution via the exponential ansatz.
This is probably beyond our present possibilities  so that
one has
 to restrict the class of operators to build an ameneable
ansatz.
 If this constructions were possible, the parameters $\alpha
_I$  would give us information about the size of the
CDW-patches in
 the true ground state. Work in this direction will be
reported elsewhere.

In this paper we have chosen a variational technique in
order to  compute the parameters $\alpha _I$ of the
exponential ansatz.  There is however an alternative way to
compute this parameters  which is called the Fokker-Planck
method \cite{js}.
 This method tries to generalize to quantum lattice
hamiltonians  the well-known fact that in quantum mechanics,
given a wave  function of exponential form $\psi (x) =
e^{-s(x)}$ one can find a  hamiltonian $H_{FP}$, called
Fokker-Planck hamiltonian,  for which $\psi (x)$ is an exact
ground state.

In the case of the ITF model, the Potts model and others the
 FP-method can be easily generalized for states of the form
(\ref{3}) due to the fact that the operators $V_I$ commute
among themselves (in a sense they play the role of the
$x$-coordinate in the  quantum mechanical case.) Then, given
the ansatz (\ref{3}) one can find the corresponding
FP-hamiltonian  and can in turn compare with the original
hamiltonian. From this
 comparison one finds the $\alpha _I 's$ in terms of the
coupling constants in $H$.

It is an interesting question to see whether the FP-method
can
 be extended to problems with  fermions. The first problem
that we encounter is that the  operators $V_I$ no longer
commute  among themselves (see Appendix C). The advantage of
the
 FP-approach is that one can introduce and do calculations
with a  large number of parameters $\alpha _I $ which serve
to introduce  much of the expected physics in the ansatz.

To summarize, we have employed a Perturbative-Variational
(PV) method to study strongly interacting fermions at
half-filling and zero temperature. To lowest order in PV  we
have stressed the similarities and differences with the
mean field theory or Hartree-Fock method. To higher order
in PV they clearly differ.
 In $1D$  our method is capable of unveiling the existence
of a critical point in the  coupling constant at
$(t/U)_c=0.7483$ as in fact occurs in  the exact solution at
a value of $0.5$. In our approach this phase transition is
described  as an  example of Bifurcation Phenomena in the
ground state energy. In $2D$  the van Hove singularity
plays an essential  role in changing the asymptotic
behaviour of the system  for large values of $t/U$. In
particular, the staggered  magnetization for large $t/U$
does not display the
 Hartree-Fock law $(t/U) e^{-2 \pi \sqrt{t/U}}$ but instead
we find the law $(t/U) e^{- \frac{\pi ^2}{3} t/U}$.
Moreover, the system does not exhibit bifurcation  phenomena
and thus we do not find a critical point separating  a CDW
state from a fermion ``liquid" state.

Many of the ideas and techniques that we propose in this
paper can in principle be applied to the
 Hubbard model in the strong coupling regime ($U>t$), which
is believed to be one of the most prominents candidate for a
microscopic description of the high-$T_c$ cuprates.

\vspace{30 pt}

{\bf Acknowledgements}
\vspace{20 pt}

Work partially supported in part by CICYT under  contracts
AEN93-0776 (M.A.M.-D.) and PB92-1092 (G.S.).

We want to thank J.G. Esteve, F. Jimenez and  M.A.
Garcia-Bach for useful discussions.

It is a pleasure to thank A. Gonzalez for sharing with  us
his access to the Alpha computer Ciruelo to make  some of
the numerical computations in this work.


\newpage

\newcommand{\Appendix}[1]{\appendix{#1}\setcounter{equation}{0}}


\appendix

\Section{The PV Formulas for Interacting Fermions}


In section 2 we have briefly explained the fundamentals of
the  perturbative-variational method and at the end of it we
have set
 up the basic equations (\ref{19})-(\ref{21}) which encode
the  essence of our method as applied to the system of
interacting  fermions. In this appendix we give the
highlights and some  details of the computations leading to
those equations.

The hamiltonian (\ref{4}) which is the subject of this study
is  split into two pieces,

\begin{equation} H = H_0 +
H_1
\label{a1}
\end{equation}

\begin{equation} H_0 = U \sum _{<i,j>} (n_i-\frac{1}{2})
(n_j-\frac{1}{2})  \label{a2}
\end{equation}

\begin{equation}
 H_1 = -t \sum _{<i,j>} (c_i^{\dagger} c_j + c_j^{\dagger }
c_i)  = t \sum _k \epsilon _k
n_k
\label{a3}
\end{equation}

\noindent The norm of the variational state

\begin{equation}
\psi (\alpha ) =  \exp ( \frac{\alpha }{2}\sum _{<i,j>}
(c_i^{\dagger} c_j + c_j^{\dagger } c_i) )
 |0> \label{a4}
\end{equation}

\noindent plays a major role in the calculation of all the
averaged  quantities and correlators of the theory. We shall
denote this norm  by $Z$ for it recalls the partition
function of an associated statistical  model. To see this,
let us compute its value.

 \begin{equation} Z = <\psi (\alpha ) | \psi (\alpha )> =
<CDW | \exp (-\alpha \sum _k \epsilon _k n_k) |
CDW>
\label{a5}
\end{equation}

\noindent To compute this partition function we note that
the variational state (\ref{a4})
 can be given the following representation in momentum space,

\begin{equation} |\psi > = \prod _k ' (u_k c_k^{\dagger } +
v_k c_{k+Q}^{\dagger }  |0>
\label{a6}
\end{equation}

\begin{equation} u_k = \frac{e^{-\frac{\alpha }{2} \epsilon
_k}}{\sqrt{2}},  v_k = \frac{e^{-\frac{\alpha }{2} \epsilon
_{k+Q}}}{\sqrt{2}} =
 \frac{e^{\frac{\alpha }{2} \epsilon _k}}{\sqrt{2}}
\label{a7}
\end{equation}

\noindent This representation allows us a readly derivation
of the norm in terms of the functions $u_k$, $v_k$ namely,
 $Z=\prod _k ' (u_k^2+v_k^2)=\prod _k' \cosh (\alpha
\epsilon _k)$,  and we have made extensive use of the
identity
 $\epsilon _k = -\epsilon _{k+Q}$. In order to perform  the
continuum limit it is convenient to introduce the free
energy $f_N$ of the associated statistical model by means of
the identity

\begin{equation} Z = <\psi (\alpha ) | \psi (\alpha )>
\equiv \exp (N f_N(\alpha ))
\label{a8}
\end{equation}

\noindent Then, from the previous considerations  we arrive
at the final expression for the free energy.

\begin{equation}
 \exp (N f_N(\alpha )) = \exp (\frac{1}{2} \sum _k \ln
(\cosh (\alpha \epsilon _k)) )
\label{a9}
\end{equation}

\begin{equation}
 f(\alpha )   \equiv \lim_{N\rightarrow \infty}  f_N(\alpha
) = \frac{1}{2}
\int _{\Omega } \frac{d^D k}{(2 \pi )^D} \ln \cosh (\alpha
\epsilon _k)
\label{a10}
\end{equation}

\noindent Interestingly enough, equation (\ref{a10})
represents the free energy of a system of free fermions
placed on a $D$-dimensional lattice as we already advanced.

\noindent The expectation value of the $H_1$ part of the
hamiltonian is easy to evaluate in terms of the asociated
free energy

\begin{equation}
\frac{\partial Z}{\partial \alpha } = -  <\psi (\alpha ) |
\sum _k \epsilon _k n_k |\psi (\alpha )>
\label{a11}
\end{equation}

\begin{equation} <H_1> = -t \frac{\partial }{\partial \alpha
} (\ln Z (\alpha ))
\label{a12}
\end{equation}

\noindent with the following result already expressed in
the continuum limit,

\begin{equation} <H_1> = -\frac{t N}{2}
\int _{\Omega } \frac{d^D k}{(2 \pi )^D} \epsilon _k \tanh
(\alpha \epsilon _k)
\label{a13}
\end{equation}


This much for the mean value of $H_1$ in the  exponential
ansatz. As for the $H_0$ part of the hamiltonian,  we need
the expectation value $<\sum _{<i,j>} n_i n_j>$  which
represents the correlation function of two density
operators.  Transforming the density operator to momentum
space

\begin{equation} n_i = \frac{1}{N} \sum _{k_1,k_2} e^{-i
(k_1-k_2) R_i}  c_{k_1}^{\dagger } c_{k_2}
\label{a14}
\end{equation}

\noindent the afore mentioned density-density operator reads,

\begin{equation}
\sum _{<i,j>} n_i n_j = \frac{1}{N} \sum _{k_1,k_2,k_3,k_4}
\delta _{k_1+k_2,k_3+k_4} \sum _{a=1}^D e^{i k_a^{13}}
c_{k_1}^{\dagger } c_{k_3} c_{k_2}^{\dagger}
c_{k_4}           \label{a15}
\end{equation}

\noindent where $k_a^{13}\equiv k_{a,1}-k_{a,3}$. At this
stage it is aparent that the way to compute this  correlator
is Wick's theorem. Thus the fermion correlator in $k$-space
is

\begin{equation} <c_{k_1}^{\dagger } c_{k_2}> = (\delta
_{k_1-k_2,0} + \delta _{k_1-k_2,Q} )
\frac{u_{k_1} u_{k_2}}{u_{k_1}^2 + v_{k_2}^2}
\label{a16}
\end{equation}

\noindent where $u_k=v_{k+Q}$. Therefore the basic
 contractions are the following two,

\begin{equation} <n_k>  =  <c_k^{\dagger } c_k>  =
\frac{e^{-\alpha \epsilon _k}}{e^{\alpha \epsilon _k} +
e^{-\alpha \epsilon _k}}
\label{a17}
\end{equation}

\begin{equation} <c_k^{\dagger } c_{k+Q}> =
\frac{1}{e^{\alpha \epsilon _k} + e^{-\alpha \epsilon
_k}}
\label{a18}
\end{equation}

The contraction in equation (\ref{a17}) represents  the
average number of particles with momentum $k$  and we notice
that it exhibits the standard Fermi-Dirac  distribution form
with the parameter $\alpha $ playing the  role of the
inverse temperature but recall that all our formalism
 is at zero temperature.

\noindent A lengthy but straightforward calculation yields

\begin{equation}
\frac{1}{N} \sum _{<i,j>} <n_i n_j> = \frac{1}{4} -
\frac{D}{4} I_0^2 -
\sum _{a=1}^{D}  \frac{1}{4}
I_a^2
\label{a19}
\end{equation}

\noindent where

\begin{equation} I_0 (\alpha ) = \frac{1}{N} \sum _k
\frac{1}{\cosh (\alpha \epsilon _k)} \rightarrow
\int _{\Omega } \frac{d^D k}{(2 \pi)^D} \frac{1}{\cosh
(\alpha \epsilon _k)}
\label{a20}
\end{equation}

\begin{equation} I_a (\alpha ) = \frac{2}{N} \sum _k
\frac{\cos (k_a)}{1 + e^{-2 \alpha \epsilon _k}}
\label{a21}
\end{equation}

\noindent As a result the density-density correlation
function turns out to be

\begin{equation}
\frac{1}{N} \sum _{<i,j>} <(n_i-\frac{1}{2})
(n_j-\frac{1}{2})> = -\frac{D}{4} (I_0^2 +
  I_1^2
)
\label{a22}
\end{equation}

Finally, the expectation value we were searching for is

\begin{equation} <H_0> = U \sum _{<i,j>} <(n_i-\frac{1}{2})
(n_j-\frac{1}{2})> = -\frac{ U D}{4} (I_0^2 +
  I_1^2
)
\label{a25}
\end{equation}

\noindent Recalling that

\begin{equation} <H_1> = - t N \frac{\partial f}{\partial
\alpha }
\label{a26}
\end{equation}

\noindent altogether the average energy is

\begin{equation} <H> = E = - t N \frac{\partial f}{\partial
\alpha }   -\frac{ U D N}{4} (I_0^2 +
  I_1^2
)
\label{a27}
\end{equation}

\noindent Using the identity $\frac{\partial f}{\partial
\alpha } = D I_1(\alpha )$  the average energy is

\begin{equation} E = - t N D I_1   -\frac{ U D N}{4} (I_0^2 +
  I_1^2
)
\label{a28}
\end{equation}

\noindent For the sake of simplicity it is convenient to
introduce the  reduced energy as follows

\begin{equation} e \equiv -\frac{ 4 E}{U N D}  = 4
\frac{t}{U} I_1   + I_0^2 +
I_1^2
\label{a29}
\end{equation}

\noindent The minimization condition
 $\frac{\partial e}{\partial \alpha } = 0$ for the reduced
energy is equivalent to that of the energy $E$ except  for
the fact that a minimum of $E$ corresponds to a  maximum of
$e$. Furthermore, we have

\begin{equation} e(t=0) = 1 \Longleftrightarrow E(t=0) =
-\frac{N U D}{4}
\label{a30}
\end{equation}

\newpage


\appendix
\setcounter{section}{1}

\Section{ The Asymptotic Formulas for Interacting Fermions
in $D=2$ Dimensions}


In section 4 we have addressed the issue on  the existence
of a critical value $(t/U)_c$ for which  a quantum phase
transition exists in two spatial dimensions.  To achieve
this goal in our PV approximation  scheme we need to
evaluate the $\alpha \rightarrow \infty$  asymptotic
behaviour of the basic integrals $I_0, I_1$ in (\ref{20}),
(\ref{20b}).  This was easily done in $D=1$ dimensions in
section 3 but it becomes much more cumbersome
 in $D=2$ dimensions and we present in this appendix  the
main details of this calculation.

Let us consider the integral $I_0$ in two dimensions as  the
paradigm of asymptotic evaluation:

\begin{equation} I_0 (\alpha ) = \int _{0}^{2 \pi} \int
_{0}^{2 \pi} \frac{dk_1 dk_2}{(2 \pi)^2}
\frac{1}{\cosh (2 \alpha (\cos k_1 + \cos
k_2))}
\label{b1}
\end{equation}

\noindent In the limit $\alpha \rightarrow \infty $ only
 the points in the neighbourhood of the region

\begin{equation}
\cos k_1 + \cos k_2 \equiv 0
\label{b2}
\end{equation}

\noindent contribute by a significant amount to the
integral (\ref{b1}). But these points (\ref{b1}) form
precisely the Fermi Surface (F.S.) of the system  at
half-filling. This surface is drawn in Figure 5. It  is a
square inscribed in the Brillouin zone with corners  at the
points $A (\pm \pi ,0)$, $B (0,\pm \pi )$.

\noindent In order to isolate this contribution from the
Fermi Surface we must distinguish two types of points  on
that curve.

\vspace{30 pt}

{\bf B1. Regular Points Contribution to $I_0$}

\vspace{30 pt}

We call a regular point of the F.S. to any point such  as
$P_r$ in Figure 5 which is not a corner. Thus we  have 4
sides of regular points each of which equally  contributes
to $I_0$. We restrict our attention to one  side of the
square, say the up-right one.

The contribution of the regular points come from a rectangle
centered at the corresponding F.S. portion of infinitesimal
width $2 \Lambda $ (see Figure 5). We must also isolate the
contribution of the corner points $A$ and $B$ so that we
introduce another infinitesimal quantity $\epsilon $ such
that
 the length of the rectangle is $\sqrt{2} \pi -2\epsilon $.

\noindent From Figure 5 it is apparent that both
infinitesima  are not independent. In fact,

\begin{equation}
\Lambda = \epsilon
\label{b3}
\end{equation}

\noindent They are equal but we keep both for the sake  of
keeping track clearly where everything comes from.

\noindent Given a regular point on the F.S., we can
parametrize  any point in the relevant rectangular region of
Figure 5  by  two parameters $k_0$, $\kappa $ as

\begin{equation}
\left\{ \begin{array}{l}     k_1 = k_0 + \kappa \\
               \\ k_2 = \pi - k_0 + \kappa
\end{array}                 \right.
\label{b4}
\end{equation}

\noindent where $k_0$ measures the position on  the F.S. and
$\kappa $ measures the distance from the F.S., that is ,

\begin{equation}
\left\{ \begin{array}{l}     k_0 \in  (\epsilon, \pi -
\epsilon) \\
               \\
\kappa \in (-\Lambda ,\Lambda )
\end{array}                 \right.
\label{b5}
\end{equation}

\noindent Thus, on each fringe the energy function $\cos k_1
+ \cos k_2$  contributes like $-2 \sin k_0 \sin \kappa $.
Substituting this into $I_0$ we have,

\[
\frac{2}{(2 \pi )^2} \int _{\epsilon}^{\pi -\epsilon } dk_0
\int _{-\Lambda }^{\Lambda } d\kappa
\frac{1}{\cosh (4 \alpha \sin k_0 \sin \kappa )}
\]

\noindent As $\Lambda \ll 1$ we can approximate
 $\sin \kappa \simeq \kappa $. Changing variables to
$x=(4 \alpha \sin k_0) \kappa $ it is possible to  perform
the $\alpha \rightarrow \infty$ limit on the  boundaries of
integration in $\kappa $ to obtain

\[
\frac{2}{(2 \pi )^2} \int _{\epsilon}^{\pi -\epsilon } dk_0
\frac{1}{4 \alpha \sin k_0}
\int _{-\infty }^{\infty } dx
\frac{1}{\cosh x}
\]

\noindent These integrals are known and  collecting the
factor
$4$ from the same number of fringes we arrive at the
leading behaviour for $I_0^{(D=2)}$ coming from the regular
points (R.P.),

\begin{equation} I_{0,RP}^{(D=2)} \sim -\frac{1}{\pi \alpha
} \ln \epsilon
\label{b6}
\end{equation}

\noindent We arrive at a $1/\alpha $ behaviour but we  must
be cautious with it for it comes with a $\ln \epsilon $
term which means that it is singular as $\epsilon
\rightarrow 0$.  This means that equation (\ref{b6}) is not
the full story about $I_0$  and the contribution from the
singular points $A$ and $B$ should  contain a singular term
which cancels out (\ref{b6}). We now check  that this is in
fact the case.

\vspace{30 pt}

{\bf B2. Singular Points Contribution (Van Hove Singularity)
to $I_0$}

\vspace{30 pt}

We have left out for the last moment the contribution from
the  corner points $A (\pm \pi ,0)$ and
$B (0, \pm \pi )$. This amounts to the integration over  the
dashed region in Figure 5. Due to the identifications  of
the Brillouin zone boundaries, we can collect the 4  dashed
regions into 2 dashed squares centered at
$A (\pi ,0)$ and $B (0, \pi )$ by gluing them together.

\noindent As a matter of fact, both dashed squares
contribute  the same amount to $I_0$ so that we concentrate
only on the  evaluation of one of them, say $A (\pi ,0)$.
Let us denote by $vH_A$  this dashed square around A. We can
parametrize any point in
 $vH_A$ by $\kappa _1$, $\kappa _2$ as

\begin{equation}
\left\{ \begin{array}{l}     k_1 = \pi + \kappa _1 \\
               \\ k_2 =  \kappa _2
\end{array}                 \right.
\label{b7}
\end{equation}

\noindent Thus, on this dashed square the energy  function
$\cos k_1 + \cos k_2$ contributes like

\begin{equation}
\cos k_1 + \cos k_2 \sim \frac{1}{2} (\kappa _1^2 - \kappa
_2^2)         \label{b8}
\end{equation}

Notice that this is a tipically van Hove-like singularity
behaviour revealing the saddle-point nature of the  singular
points A and B. Inserting (\ref{b8}) in $I_0$ we get

\[
\frac{1}{(2 \pi)^2} \int \int _{vH_A} d\kappa _1 d\kappa _2
\frac{1}{\cosh (\alpha (\kappa _1^2 - \kappa _2^2))}
\]

\noindent Now it is convenient to transform to light-cone
coordinates $\kappa _{\pm }$,

\begin{equation}
\left\{ \begin{array}{l}
\kappa _1 = \frac{1}{\sqrt{2}} ( \kappa _+ + \kappa _- ) \\
               \\
\kappa _2 =  \frac{1}{\sqrt{2}} ( \kappa _+ - \kappa _- )
\end{array}
\right.
\label{b9}
\end{equation}

\noindent and thus we arrive at

\[
\frac{4}{(2 \pi )^2} \int _0^{\epsilon} \int _0^{\epsilon }
d\kappa _+ d\kappa _-
\frac{1}{\cosh (2 \alpha \kappa _+ \kappa _-)}
\]

\noindent Moreover, this integral can be evaluated
 by changing to variables $(u,v)$ as,

\begin{equation}
\left\{ \begin{array}{l}     u = \kappa _+  \kappa _-  \\
               \\ v =  \kappa _+ / \kappa _-
\end{array}
\right.
\label{b10}
\end{equation}

\noindent Eventually, after some algebra we  obtain the
following contribution from the van Hove point A:

\begin{equation}
\frac{1}{2 \pi \alpha } \ln \epsilon - \frac{1}{2 (2 \pi
)^2}
\int _0^{\epsilon ^2} du \frac{\ln u}{\cosh (2 \alpha
u)}
\label{b11}
\end{equation}

\noindent Notice that the $\ln \epsilon $ contribution  in
(\ref{b11}) when multiplied by 2 singular points yields  the
desired cancelation of the singular behaviour obtained  in
(\ref{b6}) from the regular points.

In all, we are left with a finite contribution that
 can be extracted from (\ref{b11}).  Therefore, the leading
order behaviour of $I_0 (\alpha )$
 as $\alpha \rightarrow \infty $ is given by

\begin{equation} I_0^{(D=2)}(\alpha ) \longrightarrow
\frac{\ln \alpha }{2 \pi \alpha }
\label{b12}
\end{equation}

The remarkable fact about this asymptotic behaviour is
 the presence of the $\ln \alpha $ term in addition to  the
$1/\alpha $ term already found in $D=1$ dimensions.  This in
turn is the distintive characteristic feature of the
$D=2$ dimensional problem.

The technique outlined above to isolate the leading
contribution of $I_0(\alpha )$ can also be applied to
$dI_0(\alpha )/d\alpha $ and $dI_1(\alpha )/d\alpha $ (in
fact $I_1(\alpha )$  is easier to handle and does not
requires that technique).  In  Table 2 we summarize the
final results as well as the  intermediate results showing
explicitly that the cancelation
 of the singular $\ln \epsilon $ terms also occurs for
$dI_0(\alpha )/d\alpha $ and $dI_1(\alpha )/d\alpha $.

\newpage


\appendix
\setcounter{section}{2}

\Section{ Fokker-Planck Hamiltonians for Fermions}


In this appendix we want to describe in more  detail the
problems arising when trying to apply
 the FP-method of reference \cite{js} to fermions.  The
question we want to answer is whether one
 can construct a hamiltonian for which the ansatz
 $\psi (\alpha )$ (\ref{10}) becomes an exact eigenstate.
 Following \cite{js} we should act with $H_0$ on the state
 (\ref{10}) and using the anticommutation relations of the
operators $n_j-\frac{1}{2}$, $c_j$ and $c_j^{\dagger }$

\[ H_0 | \psi (\alpha ) >  = -\frac{U}{4} \sum _{<i,j>}
\exp (-\frac{\alpha }{2} \sum _{<i,k>, k\neq j} T_{ik}
-\frac{\alpha }{2} \sum _{<j,k>, k\neq i }T_{jk} +
\mbox{rest unchanged}) |CDW>
\]
\begin{equation} = -\frac{U}{4} \sum _{<i,j>}
\exp (-\alpha  \sum _{<i,k>, k\neq j} T_{ik} -
\alpha  \sum _{<j,k>, k\neq i }T_{jk} + \frac{\alpha }{2}
\sum _{<l,m>}T_{lm} )
|CDW>
\label{c1}
\end{equation}

\noindent If the operators $\{ T_{ij} \}$ commute  among
themselves we could take the term
$\frac{\alpha }{2} \sum _{<l,m>}T_{lm} $ to the right  of
the exponential recovering in that fashion the  state $\psi
(\alpha )$. This is precisely what happens
 in the cases described in reference \cite{js}.  Here
however the operators do not commute and this
 complicates matters. We can apply the
Baker-Campbell-Hausdorf formula to the last exponential  in
(\ref{c1}). The commutator between the  sum of the first and
second sums with the third gives
 operators which acting on $|CDW \rangle $ vanishes. Hence
to order $\alpha ^2$ equation (\ref{c1}) becomes

\begin{equation} H_0 | \psi (\alpha ) >  \simeq -\frac{U}{4}
\sum _{<i,j>}
\exp (-\alpha  \sum _{<i,k>, k\neq j} T_{ik}  -
\alpha  \sum _{<j,k>, k\neq i }T_{jk} ) |\psi (\alpha
)>                \label{c2}
\end{equation}

\noindent So that one gets up to order $\alpha ^2$:

\begin{equation} H_{FP} | \psi (\alpha ) >  =
0              \label{c3}
\end{equation}

with

\begin{equation} H_{FP}   =  U \sum _{<i,j>} \{ (n_i -
\frac{1}{2}) (n_j - \frac{1}{2})  +
\exp (-\alpha  \sum _{<i,k>, k\neq j} T_{ik}  -
\alpha  \sum _{<j,k>, k\neq i }T_{jk} )
\}                                       \label{c4}
\end{equation}

\noindent Expanding the exponential in powers  of $\alpha $
we observe that the terms linear in $\alpha $  correspond
precisely to $U_1$ with the correct relationship
 between $\alpha $ and $t/U$ (\ref{23}).
 In fact, from these type of comparison is how one derive
 the $\alpha $-parameters as functions of the coupling
constants. Now, to order $\alpha ^2$ we also find the
operators appearing in $U_2$, but there are some more
operators which acting on the CDW-vacuum give nothing.
 The solution of the PV-equations is by no means unique.

\noindent Higher order parameters in $\alpha $ give some
 new terms but also others that are already contained in
$U_1$ and $U_2$. This means in particular that the relation
between $t/U$ and  $\alpha$ is rather non-linear,
 $t/U = g(\alpha ) = (D-\frac{1}{2}) \alpha + \ldots $.  In
order to improve the ansatz one should add to
 the exponential of the ansatz the unwanted operators that
appear in the FP-hamiltonian (we mean those operators which
do not appear in the original ansatz). This program has been
successfully implemented for Ising-like models but it
remains  to be seen whether  it can also be extended for
fermions.


%
%
\def\baselinestretch{1.5}
\noindent

\newpage
\section*{Table captions}

 {\bf Table 1 :} Reduced ground state energy $e(t/U=1/2)$
values
 for the isotropic case as compared to several methods in
D=1 dimensions.

{\bf Table 2 :} Asymptotic leading order behaviour for the
basic  integrals in $D=2$ dimensions showing the cancelation
of the  singular $\ln \epsilon $ terms between the
contributions of the
 regular points on the Fermi Surface and the van Hove
singularities.

\newpage
\section*{Figure captions}
\noindent

 {\bf Figures 1 a) - d) :} Reduced energy $e(\alpha ;t/U)$
in $1D$ eq. (\ref{19b})
 for several values of the coupling constant $t/U$ showing
 the bifurcation phenomena responsible for the existence  of
a critical point $(t/U)_c$ in the PV approach.

 {\bf Figure 2 :}  The coupling constant $t/U$ as a function
of the PV parameter $\alpha $ for $1D$  (solid line) showing
the existence of a critical point $(t/U)_c$  and $2D$ (grey
line) showing the $\ln \alpha $ behaviour for large $\alpha
$.

 {\bf Figure 3 :} Staggered magnetization versus
 the coupling constant ($t/U$): exact $1D$ result (solid
line),
 PV $1D$ result (strong solid line) and PV $2D$ result (grey
line).

 {\bf Figures 4 :}  Reduced energy versus the coupling
constant
 ($t/U$): exact $1D$ result (black line), PV $1D$ result
 (black-grey solid line) and PV $2D$ result (grey line).

 {\bf Figure 5 :} Brillouin zone for $2D$ fermions at
half-filling  showing the Fermi Surface formed of regular
points $P_r$ and  van Hove singular points $A (\pi ,0)$ and
$B (0,\pi)$.
 $\Lambda =\epsilon $ are the cut-offs.

\newpage

\begin{table}[p]
\centering
\begin{tabular}{|l|c|c|}  \hline \hline
  $\mbox{Method}$    & \mbox{Reference} &   $e(t/U=1/2)$
\mbox{Isotropic case}  \\  \hline \hline
 Neel   &      &  $  1 $  \\ \hline Free Fermi Sea    &
\cite{rodri} &  $ 1.6766 $  \\  \hline
 Present PV solution &    &  $  1.7232$  \\  \hline
 Jordan-Wigner transformed UHF solution     &  \cite{rr}
&  $1.7292$     \\ \hline RVB ansatz $(M=3)$      &
\cite{gbach}  &   $1.7580$   \\ \hline Exact solution  &
\cite{hulthen}    &  $1.7726$    \\ \hline \hline
\end{tabular}
\caption{Reduced ground state energy $e(t/U=1/2)$ values for
the isotropic case as compared to several methods in D=1
dimensions.}
\end{table}

\begin{table}[p]
\centering
\begin{tabular}{|l|c|c|c|}  \hline \hline
  $D=2$ \mbox{ Integrals}    & \mbox{ $\alpha \rightarrow
\infty$ Behaviour} &  \mbox{R. P. Contribution } & \mbox{Van
Hove Contribution } \\  \hline \hline
 $I_0^{(D=2)} (\alpha )$    &  $\frac{\ln \alpha }{2 \pi
\alpha } $  &  $ -\frac{1}{ \pi \alpha } \ln \epsilon
$       & $\frac{1}{ \pi \alpha } \ln \epsilon + \frac{\ln
\alpha }{2 \pi \alpha } $ \\ \hline
$\frac{dI_0^{(D=2)}}{d\alpha }(\alpha ) $   &  $-\frac{\ln
\alpha }{2 \pi \alpha ^2} $&
$ \frac{1}{ \pi \alpha ^2} \ln \epsilon  $  & $ -\frac{1}{
\pi \alpha ^2} \ln \epsilon  -\frac{\ln \alpha }{2 \pi
\alpha ^2} $ \\  \hline
 $I_1^{(D=2)} (\alpha )$ & $(\frac{2}{\pi })^2$   &    & \\
\hline
 $\frac{dI_1^{(D=2)}}{d\alpha }(\alpha ) $   & $\frac{\ln
\alpha }{24 \alpha ^3} $  &
 $-\frac{1}{ 24 \alpha ^3} \ln \epsilon$  &
$\frac{1}{ 24 \alpha ^3} \ln \epsilon + \frac{\ln \alpha
}{24 \alpha ^3}$    \\ \hline \hline
\end{tabular}
\caption{Asymptotic leading order behaviour for the basic
 integrals in $D=2$ dimensions showing the cancelation  of
the singular $\ln \epsilon $ terms between the contributions
of the regular points on the Fermi Surface and the van Hove
singularities.}
\end{table}

\end{document}